\documentclass[]{aa}
\usepackage{graphicx}
\usepackage{epsfig}
\def\lappr{\lower 3pt\hbox{$\buildrel < \over \sim\;$}}
\def\llappr{\lower 3pt\hbox{$\buildrel > \over \sim\;$}}
\begin{document}
   \title{XMM-Newton observations expose 
AGN in apparently normal galaxies}

   \author{P. Severgnini
          \inst{1}
	  \and
	 A. Caccianiga\inst{1}
	 \and
	 V. Braito\inst{1,2}
	 \and
	 R. Della Ceca\inst{1}
	 \and
	 T. Maccacaro\inst{1}
	 \and
	 A. Wolter\inst{1}
         \and
	 K. Sekiguchi\inst{3}
	 \and
	 T. Sasaki\inst{3}
	 \and 
	 M. Yoshida\inst{4}
	 \and 
	 M. Akiyama\inst{3}
	 \and
	 M. G. Watson\inst{5}
	 \and
	 X. Barcons\inst{6}
	 \and
	 F.J. Carrera\inst{6}
	 \and
	 W.Pietsch\inst{7}
	 \and
	 N. A. Webb\inst{8}
	  }

   \offprints{P. Severgnini, e-mail: paola@brera.mi.astro.it}

   \institute{Osservatorio Astronomico di Brera, Via Brera 28, I-20121, 
     Milano, Italy\\
     \email{paola, caccia, braito, rdc, tommaso, anna@brera.mi.astro.it}
   \and
     Dipartimento di Astronomia, Universit\`a di  Padova, 
     Vicolo dell'Osservatorio 2, 
     I-35122, Padova, Italy
   \and
     Subaru Telescope, National Astronomical Observatory of Japan\\
     \email{kaz, sasaki, akiyama@subaru.naoj.org}
   \and
     Okayama Astronomical Observatory, NAOJ\\
     \email{yoshida@oao.nao.ac.jp}
   \and
     X-ray Astronomy Group, Department of Physics and Astronomy, 
     Leicester University, Leicester LE1 7RH, UK\\
     \email{mgw@star.le.ac.uk}
   \and
     Instituto de Fisica de Cantabria (CSIC-UC), Avenida de los Castros, 
     39005 Santander, Spain\\
     \email{barcons, carreraf@ifca.unican.es}
   \and
     Max-Planck-Institut fur extraterrestrische Physik, 
     85741 Garching, Germany\\
     \email{wnp@mpe.mpg.de}
   \and
    Centre d'Etude Spatiale des Rayonnements, 9 avenue du Colonel Roche, 
    31028 Toulouse Cedex 04, France\\
     \email{webb@cesr.fr}
          }

   \date{Received ...; accepted ...}

   \abstract{
We have performed a detailed analysis of 3 optically normal galaxies extracted
from the {\it XMM Bright Serendipitous Source Sample}.  Thanks to the good
statistics of the {\it XMM-Newton} data, we have unveiled the presence of an
AGN in all of them. In particular, we detect both X--ray obscured
(N$_H>$10$^{22}$ cm$^{-2}$) and unobscured (N$_H<$10$^{22}$ cm$^{-2}$) AGNs with
intrinsic 2--10 keV luminosities in the range between 10$^{42}$ -- 10$^{43}$
erg s$^{-1}$. We find that the X--ray and optical properties of the sources
discussed here could be explained assuming a standard AGN hosted by galaxies
with magnitudes M$_R<$M$^*$, taking properly into account the absorption
associated with the AGN, the optical faintness of the nuclear emission with
respect to the host galaxy, and the inadequate set--up and atmospheric
conditions during the optical spectroscopic observations. Our new spectroscopic
observations have revealed the expected AGN features also in the optical band.
These results clearly show that optical spectroscopy sometimes can be
inefficient in revealing the presence of an AGN, which instead is clearly
found from an X--ray spectroscopic investigation. This remarks the importance
of being careful in proposing the identification of X--ray sources (especially
at faint fluxes) when only low quality optical spectra are in hand. This is
particularly important for faint surveys (such as those with {\sl XMM-Newton}
and {\sl Chandra}), in which optically dull but X-ray active objects are being
found in sizeable numbers.

   }

	\titlerunning{AGN posing as optically normal galaxies}
   \maketitle
%
%________________________________________________________________

\section{Introduction}
The existence of an intriguing population of galaxies with X--ray properties 
suggesting the presence of an AGN, but without any obvious 
sign of activity in their
optical spectra, has been claimed some 20 years ago
from the analysis of observations taken with the
{\it Einstein} Observatory (e.g. Elvis et al. \cite{Elvis81}, 
Maccacaro et al. \cite{Maccacaro1}). 
This discovery has been subsequently supported by  ROSAT data
(e.g. Griffiths et al. \cite{Griffiths}; 
Tananbaum et al. \cite{Tananbaum}; Pietsch et al. \cite{Pietsch}; 
Worrall et al. \cite{Worrall}; 
Lehmann et al. \cite{Lehmann}, \cite{Lehmann1}) 
and more recently also by {\it Chandra} 
(Fiore et al. \cite{Fiore}; Barger et al. \cite{Barger1}, \cite{Barger2}, 
\cite{Barger3}) and
{\it XMM-Newton} (Comastri et al. \cite{Comastri1}, 
\cite{Comastri2}, \cite{Comastri3}) observations.
These sources, characterized by high X--ray luminosities
(L$_X\llappr$5x10$^{41}$ erg s$^{-1}$) and X--ray--to--optical flux
ratios similar to those of AGN (-1$\lappr$Log($\frac{F_x}{F_{opt}})\lappr$1), 
have been named  in a variety of ways: {\it Optically Dull Galaxies}
(Elvis et al. \cite{Elvis81}); 
{\it Passive Galaxies} (Griffiths et al.\cite{Griffiths}) and more recently 
{\it X-ray Bright Optically Normal Galaxies} (XBONG, 
Comastri et al. \cite{Comastri2}). 
If an AGN is really present in their nuclei,
the lack of evident optical emission lines has to be justified.
Different  scenarios have been proposed so far: 
a) the AGN is a BL Lac object; b) the nuclear activity is outshone  by 
the stellar continuum; c) the lines could not be efficiently produced
or they could be  absorbed by material beyond the NLR.
Although the broad--band properties of one XBONG 
in the HELLAS2XMM survey (Brusa et al. \cite{Brusa}) are consistent with 
the presence of a BL Lac object, which are usually X--ray unobscured sources, 
most of the optically dull galaxies seem 
to be obscured in the X--ray domain (Mainieri et al. \cite{Mainieri}).
This last finding is 
mainly based on the hardness ratio
analysis (Barger et al. \cite{Barger1}, \cite{Barger2}, 
\cite{Barger3}, Comastri et al. \cite{Comastri3}). 
Recently, Moran et al. 
(\cite{Moran}) have obtained integrated (i.e. nucleus+host galaxy) 
optical spectra of a sample of nearby 
Seyfert~2 galaxies obscured in the X--ray band in order to
simulate ground-based spectroscopic observations of the more distant  
X--ray optically
dull galaxies under investigation in deep {\it Chandra} surveys. 
They find that, due to the limitations of optical spectroscopic 
observations (e.g. wide slits, low resolution, low signal--to--noise, 
inadequate spectral coverage), optically type~2, X--ray obscured AGN could be easily
undetected in the optical band. In this context, 
Comastri et al. (\cite{Comastri3})
have recently shown that the distribution of X--ray--to--optical flux
ratio of XBONG sources could be well reproduced assuming that
the underlying SED of the putative AGN is that of an X--ray 
Compton--thick AGN.
In spite of this, the nature of this kind of sources is still far from 
being completely understood. Indeed, in the majority of the cases, 
the poor X--ray statistics obtained so far on this class of sources
has not allowed us to unambiguously establish the presence 
of an AGN nor to study its nature. 

In this paper we present new {\it XMM-Newton} and optical data for
3 sources belonging to the  {\it XMM Bright Serendipitous Sample} 
(XMM-BSS hereafter, Della Ceca et al. \cite{DellaCeca_estec};
Della Ceca \cite{DellaCeca_como}). 
On the basis of the optical spectra available from the 
literature or taken as a part of 
the AXIS\footnote{http://www.ifca.unican.es/$\sim$xray/AXIS/} 
project (Barcons et al. \cite{Barcons}, \cite{Barcons1}),
these three XMM--BSS sources were
classified, at first glance, as normal galaxies.
The same sources should be instead classified as AGN candidates
on the basis of their X--ray luminosities and 
their X--ray--to--optical flux ratios (see Sect.~2).
With the aim of investigating the true nature of these three objects
we have studied, thanks to the good X--ray statistics 
of the {\it XMM-Newton} data, their X--ray spectral properties (see Sect.~3).
This analysis has allowed us to develop a simple
model (described in Sect.~4.1) which, starting from the X--ray properties
of the AGN present in these sources, has allowed us to explain: 
a) why the nuclear emission is undetected
in the optical spectra already available and b) what kind of optical
observations are needed to unveil the AGN also in the optical domain.
Subsequent spectroscopic observations performed with a more  suitable
observational set--up (i.e. narrower slits or larger spectral coverage) 
have allowed us to unveil the AGN features in the optical band.
The main results obtained for each source are discussed in Sect.~4.2, 4.3
and the conclusions are reported in Sect.~5.
Throughout this paper we assume 
H$_0$=65 km s$^{-1}$ Mpc$^{-1}$ and $\Omega_M$=0.3, $\Omega_{\Lambda}$=0.7.

%__________________________________________________________________

\section{XMM-BSS Optical Dull Galaxies: broad-band properties}

The XMM-BSS sample is an on--going project carried out by 
the {\it XMM Survey Science Centre} (XMM--SSC, Watson et al. \cite{Watson}) 
with the aim of complementing the results obtained by the deep
{\it Chandra} surveys (CDF-N, Brandt et al. \cite{Brandt}; 
CDF-S, Rosati et al. \cite{Rosati}, Giacconi et al. \cite{Giacconi}) 
and by the deep (Lockman Hole, Hasinger et al. \cite{Hasinger}) 
and medium--deep {\it XMM-Newton} surveys
(AXIS, Barcons et al. \cite{Barcons}, \cite{Barcons1}; 
HELLAS2XMM, Baldi et al. \cite{Baldi}). It is planned as a large 
($\sim$1000 sources), high Galactic latitude ($|b_{II}|>$20$^{o}$) 
sample of bright (F$_X>$10$^{-13}$ cgs) 
serendipitous {\it XMM-Newton}  sources.
The sample definition and selection criteria 
are described in Della Ceca et al. (\cite{DellaCeca_estec}, 
\cite{DellaCeca_como}).

\begin{table*}[t!]
\caption{The 3 {\it X-ray Bright Optically Normal Galaxies} discussed in this 
paper.}
\label{optical_xray data}
\small
\begin{tabular}{cccccccc}
\hline 
\hline 
% & & \\  
\# &  XMMJ & R &  z & F$_{0.5-4.5\rm keV}$ & F$_{2-10\rm keV}$ & 
L$_{2-10\rm keV}$ & Log (F$_{2-10\rm keV}$/F$_{opt}$)$^a$\\
   &       & (mag) & &  (10$^{-13}$ cgs) & (10$^{-13}$ cgs) & (10$^{42}$ cgs) & \\
\hline 
\hline 
% & & \\  
1 & 021822.3-050615.7 & 14.8$^1$ & 0.044$^2$  & 0.39 & 3.01$^3$  & 1.3  & -1.23\\
  & & \\  
2 & 031859.2-441627.6 & 16.7$^4$ & 0.1395$^4$ & 1.18 & 1.63 & 8.8 & -0.64\\
  & & \\
3 & 075117.9+180856.1 & 18.3$^5$ & 0.255$^2$  & 1.32 & 1.64 & 30.6 & -0.03 \\   
\hline
\hline   
\end{tabular}
\\
~~~~\\
%$^a$ MOS flux (corrected for the Galactic absorption, but not 
%corrected for the intrinsic absorption) in units of 10$^{-13}$ erg s$^{-2}$ cm$^{-2}$ \\
%$^b$ MOS luminosity (corrected for the Galactic absorption, but not 
%corrected for the intrinsic absorption) in units of 10$^{42}$ erg s$^{-2}$ \\
$^a$ F$_{opt}$ is the integrated flux in the Cousin R band system. 
More specifically:
$F_{opt}=FWHM \cdot (f_{\lambda\rm eff}^{Vega} \cdot 10^{(-0.4 \cdot m_R)}$). \\ 
$^1$ The magnitude reported for this source is in the Sloan r' filter. 
In order to calculate the
$F_{opt}$ for this source the r' magnitude has been transformed in
R$_{Cousin}$ magnitude using (r'- R$_{Cousin}$)=0.25 
(Fukugita et al. \cite{Fukugita}).\\
$^2$ AXIS project (Barcons et al. \cite{Barcons}, \cite{Barcons1}).\\
$^3$ This source belongs only to the XMM--BSS sample defined in the 4.5--7.5 keV
hard band. The flux of the source in the hard band is:
F$_{4.5-7.5\rm keV}$=2.43x10$^{-13}$ erg s$^{-2}$ cm$^{-2}$.\\
$^4$ From the ``ESO Nearby Abell Cluster Survey'' catalogue 
(Katgert et al. \cite{Katgert}, 
ftp://adc.gsfc.nasa.gov/pub/adc/archives/journal\_tables/A+AS/129/399/).\\
$^5$ USNO magnitude.\\
\end{table*}
\begin{figure*}[]
{\centerline{\epsfig{file=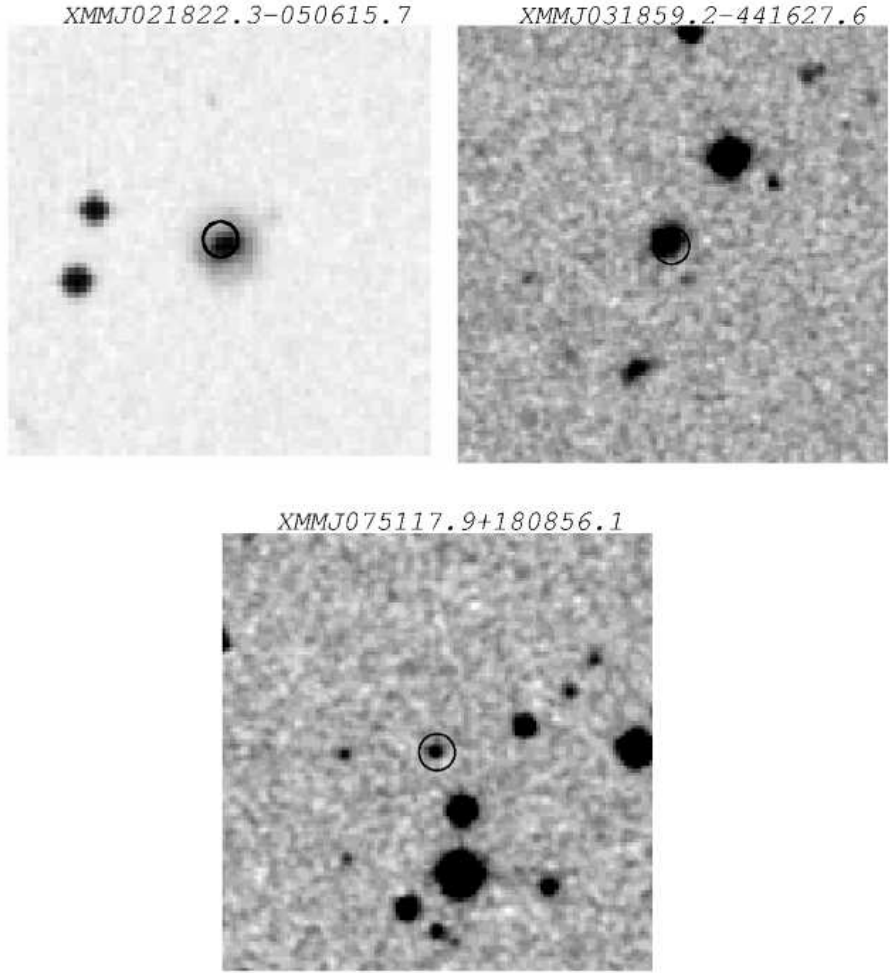, width=12cm, height=13.5cm, angle=0}}} 
%{\centerline{\epsfig{file=pippo.ps, width=12cm, height=13.5cm, angle=0}}} 
\caption{2\arcmin x 2\arcmin~ POSS II images (POSS I for XMMJ021822.3-050615.7) 
of the sources considered in this paper. North is up, east to the right.
In order to clearly mark the optical counterpart of the 
{\it XMM-Newton} source a  circle of 5\arcsec radius is shown in each image.
It is worth noting that 
the actual positional error of XMM sources
at the bright fluxes considered here is much smaller ($\sim$3\arcsec).}
\label{}
\end{figure*}

The relevant information on the optical and X--ray properties of the 3
sources discussed here are listed in Table~1.
The first column gives an identification number that, 
for simplicity, will be used in this paper in place of the full 
{\it XMM-Newton} name reported in column 2.
Columns 3 -- 8 give: R magnitudes, redshifts, X--ray fluxes and
luminosities corrected only for the Galactic absorption, and 
X--ray--to--optical flux ratios. 
Since the MOS calibration files are usually the best ones, 
the fluxes and luminosities used in this paper have been computed 
on the basis of the MOS data.

The optical finding charts are shown in Fig.~1 and more 
information on the individual sources 
are reported in the following sections.
Sources \#1 and \#3 have been classified as normal galaxies on the
basis of the optical spectra taken by the AXIS project 
(Barcons et al. \cite{Barcons}, \cite{Barcons1}), while
the optical classification of source \#2 is taken from
Katgert et al. (\cite{Katgert}).
Only for source \#1 the optical spectrum available covers the region 
where the H$\alpha$ line is expected ($\sim$6852 \AA), while for the 
remaining two sources the H$\alpha$ region is not sampled.

The X--ray properties of these three sources 
strongly suggest the presence of AGN activity: they have an
X--ray luminosity L$_{2-10\rm keV}>$10$^{42}$ erg s$^{-1}$ (see Table~1) 
and an X--ray--to--optical flux ratio similar to the typical value of
luminous AGN (-1$\lappr$Log($\frac{F_x}{F_{opt}}$)$\lappr$1; 
Maccacaro et al. \cite{Maccacaro}; 
Schmidt et al. \cite{Schmidt}; Akiyama et al. \cite{Akiyama}; 
Lehmann et al. \cite{Lehmann}).

All the 3 sources presented here lie within the area
covered by the {\it Two Micron All Sky Survey} (2MASS) and
they have been detected 
in all near-infrared (NIR) bands (J, H, K). 
Only for source \#2 the NIR 
magnitudes are not reported in the 2MASS cataloge yet. The 
$R-K$ colors of source \#1 and \#3 (2.8 and 3.0 respectively) 
are in agreement with those expected 
for an early--type galaxy (e.g. Mannucci et al. \cite{Mannucci}).

For source~\#1 radio information is available from the {\it 
Very Large Array} (VLA) survey in the Subaru field region.
The radio flux density is 325$\pm$33 $\mu$Jy at 1.4 GHz 
(Simpson et al. private communication), 
giving a power of 3x10$^{21}$ W/Hz,
typical of a radio-quiet AGN.
Source \#2 is not covered neither by the NVSS 
({\it NRAO VLA Sky Survey}, Condon et al. \cite{Condon}) 
nor by the  FIRST ({\it Faint Images of the Radio Sky at Twenty-cm}, 
Becker et al. \cite{Becker}) survey, while 
source \#3, that lies within the area covered by the FIRST
survey, does not have a radio counterpart  within 20''
radius (5$\sigma$ upper limit on the 1.4~GHz flux density of about 1.2 mJy).

%__________________________________________________________________

\section{XMM-Newton data}
In this section we present the XMM-Newton data.
The data have been processed using the SAS 
({\it Science Analysis System}) version 5.3.3. 
Events files released from the standard pipeline have been 
filtered for high background time intervals. We have used the latest 
calibration files released by the EPIC 
team to create new response matrices that include also the
correction for the effective area at the source position in the
detector. The relevant information about the
{\it XMM-Newton} observations are reported in Table~2.
In particular the table lists: 
source identification numbers (col. 1),
observation IDs (col. 2), exposure times after  
removing time intervals affected by background flares 
and taking into account all the instruments used (col. 3), 
net counts from all the instruments used (col. 4), detectors 
used for the scientific analysis (col. 5).
\begin{table*}[t!]
%\begin{center}
\caption{XMM-Newton observations}
\small
\label{obs_log}
\begin{tabular}{ccccc}
\hline \hline   
%       &       &     &  \\
Source & Observation ID & Exposure  Time & Net Counts  & Notes\\
       &               & [ks]  &       &      \\
%       &              &           &     &  \\
\hline
\hline 
\#1    & 0112370101, 0112371001 & 281,0 & 1482$\pm$45 & MOS1+MOS2+PN \\
       &                         &       &      &              \\
\#2    & 0105660101              & 39,7  & 305$\pm$22 & MOS2+PN      \\ 
       &                         &       &      & NO MOS1: source too close to a gap \\
       &                         &       &      &               \\
\#3    & 0111100301              & 57,2  & 913$\pm$46 & MOS1+MOS2 \\
       &                         &       &      & NO PN: timing mode \\
\hline 
\hline
\end{tabular}
%\end{center}
\end{table*}

\subsection{Spectral analysis}
At the spatial resolution of {\it XMM-Newton} EPIC instruments, 
the 3 observed sources appear point--like. 
The X--ray spectra have been extracted using circular regions of 
appropriate radius (from 25\arcsec~to 30\arcsec). 
Background spectra have been extracted from larger
(from 50\arcsec~to 60\arcsec)
source--free circular regions 
close to the object. 
In order to improve the statistics, 
MOS1 and MOS2 data obtained with 
the same filter have been combined together and, finally, 
MOS and PN spectra  have been binned in order 
to have at least 20 counts per energy channel.

The spectral analysis described in the following has been  performed 
using XSPEC 11.0.1.
Afterwards, unless otherwise stated, errors are given at the 90\%
confidence level for one interesting parameter ($\Delta\chi^2$=2.71).
For each source, the MOS and PN spectra have been fitted simultaneously  
in the 0.5--10 keV band, 
leaving free the relative normalizations.
In the fitting procedure, the appropriate Galactic hydrogen column density
along the line of sight has been taken into account 
(Dickey \& Lockman \cite{Dickey}). 

Two spectra (\#2 and \#3) are well described by
a single absorbed power--law model. A pure thermal component is rejected 
for all the sources at more than 97\% confidence level and the addition 
of a thermal component to the power--law model is not statistically required.

A good fit for source \#1 is obtained with a two component model: 
a ``leaky absorbed power-law continuum\footnote{
Consists of an absorbed plus an
unabsorbed power-law with the same photon index.}'' plus a cold Iron 6.4 keV 
emission line; the addition of the latter
improves the  fit at more than 99\% confidence level.
A more detailed description of the X--ray properties of this source
will be reported in a forthcoming paper by Watson et al. (in preparation).
For all the 3 sources, the relevant best fit parameters, 
quoted in the rest frame, are 
summarized in Table~3 along with the unabsorbed luminosities. 
The best fit unfolded spectra and residuals are shown in Figures~2 to 4.

\begin{table*}
\caption{Best fit parameters to XMM-Newton data}
\label{resultt}
\begin{center}
\scriptsize
\begin{tabular}{lrcccccccc}
\hline \hline
%~~~~~~~~~\\
%\multicolumn{6}{l}{\it Model: Leaky absorbed power-law continuum} \\
{\it Model: Leaky absorbed }  &$N_{\rm H\rm (GAL)}$ & $\Gamma$ & $N_{\rm H}$ & $E_{\rm Fe-K\alpha}$ & $EW_{\rm Fe-K\alpha}$ & $\chi^2/dof$ & L$_{2-10\rm keV}$$^a$ \\
{\it power-law continua+ Gaussian Line} & $[10^{20}cm^{-2}]$ && $[10^{22}cm^{-2}]$ & [keV]      & [eV] &  &  [10$^{42}$ erg s$^{-2}$] \\
%~~~~~~~~~\\
%\hline
\hline
~~~~~~~~~\\
Source\#1 & 2.47  & 1.66$\pm0.30$ & 20.54$\pm0.40$ & 6.36$\pm0.07$& 300$\pm$120 & 89.5/63 & 5.6\\
~~~~~~~~~\\
\hline
\hline
%~~~~~~~~~\\
{\it Model: Single absorbed power-law} & $N_{\rm H\rm (GAL)}$ & $\Gamma$ & $N_{\rm H}$ & $\chi^2/dof$ & L$_{2-10\rm keV}$$^a$ \\
 &$[10^{20}cm^{-2}]$ && $[10^{22}cm^{-2}]$ &  & [10$^{42}$ erg s$^{-2}$] \\
%~~~~~~~~~\\
%\hline
\hline
~~~~~~~~~\\
Source\#2 & 2.61 & 1.72$\pm0.40$ & 0.39$\pm0.27$   & 11/14 & 9.1 \\
%~~~~~~~~~\\				       
Source\#3 & 4.11 & 1.58$\pm0.16$ & 0.11$\pm0.07$   & 36/42 & 31.7 \\
~~~~~~~~~\\				       
\hline
\hline
\end{tabular}
\end{center}
$^a$ MOS unabsorbed luminosity. \\
\end{table*}

In summary, we find that 
all the spectra are well described by a power--law model.
A thermal component is not required by the fit, in agreement with the
point--like appearance of the X--ray emission and with
the lack of evident clusters/groups in the optical images.
These facts, combined with the high intrinsic X--ray luminosities
of these objects, clearly suggest the presence of an AGN.
A cold 6.4 keV Iron emission line component is required only
for source \#1. The equivalent width ($EW_{\rm Fe-K\alpha}\sim$300 eV) 
of this line 
is typical of a Compton-thin AGN (Bassani et al. \cite{Bassani}). 
For the remaining two sources the 
upper limits on the $EW_{\rm Fe-K\alpha}$ are at least a factor 2 lower 
(at the 90\% confidence level) than the typical value expected for
a Compton-thick AGN ($\sim$ 1 keV; Bassani et al. \cite{Bassani}).
\begin{figure}[]
{\centerline{\epsfig{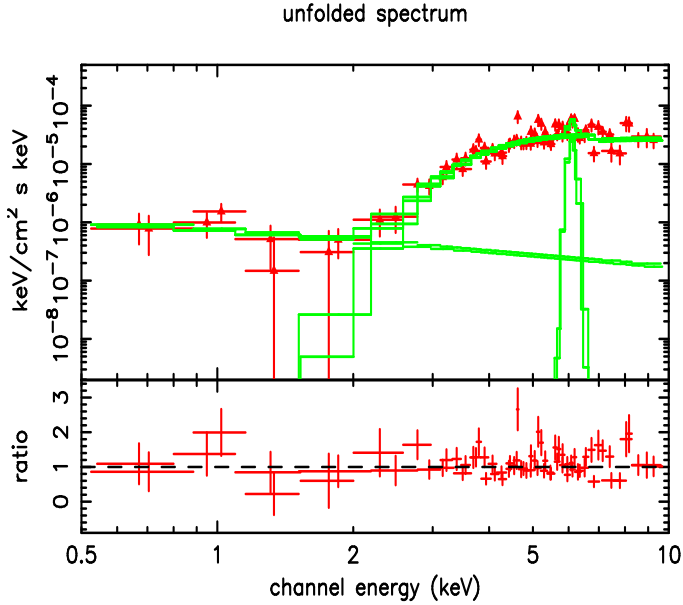}}}
\caption{ XMMJ021822.3-050615.7 (source~\#1) -- {\it Upper panel:} 
PN+MOS spectrum in energy units (unfolded spectrum, solid points) and
best--fit model (continuous line).
{\it Lower panel:} Ratio between data and the best--fit model values as a 
function of energy.}
\label{}
\end{figure}
\begin{figure}[]
%\vskip -0.85truecm
{\centerline{\epsfig{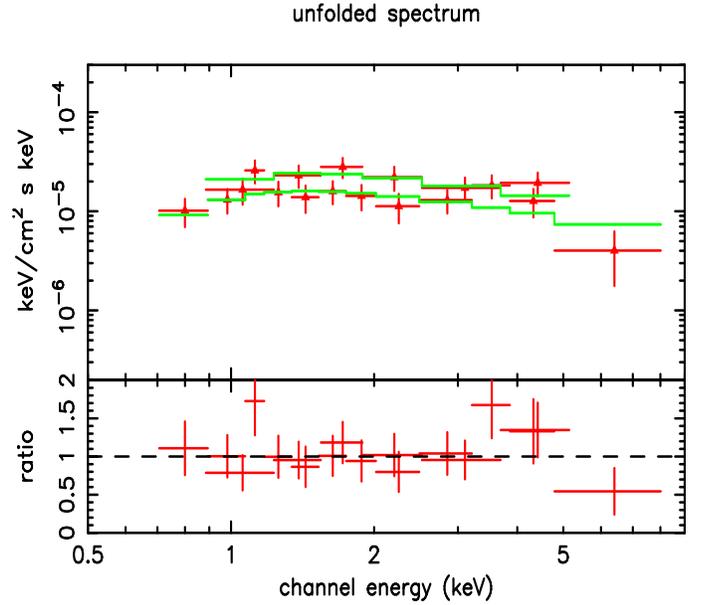}}}
\caption{ XMMJ031859.2-441627.6 (source~\#2) -- {\it Upper panel:}
PN+MOS spectrum in energy units (unfolded spectrum, solid points) and 
best--fit model (continuous lines). 
{\it Lower panel:} Ratio between data and the best--fit model values as a 
function of energy.}
\label{}
\end{figure}

\begin{figure}[]
{\centerline{\epsfig{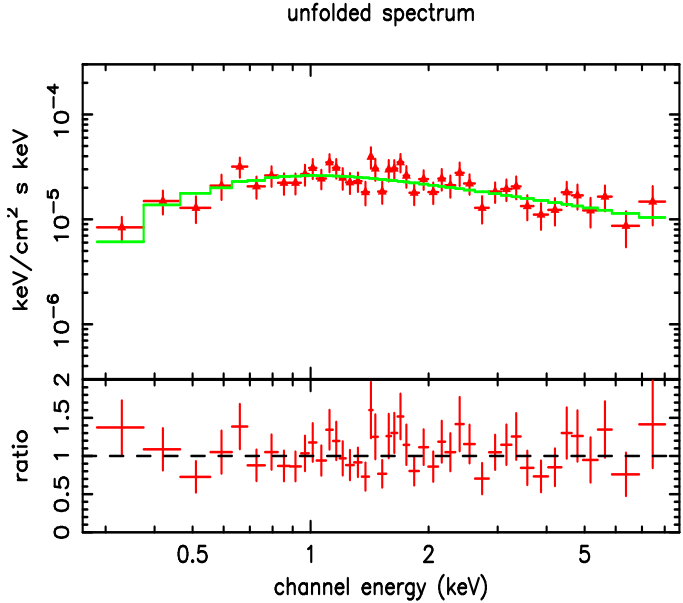}}}
\caption{ XMMJ075117.91+180856.11 (source~\#3) -- 
{\it Upper panel:} MOS spectrum in energy units (unfolded spectrum, solid point) 
and best--fit model (continuous line).
{\it Lower panel:} Ratio between data and the best--fit model values as a 
function of energy.}
\label{}
\end{figure}

\section{Discussion}

\subsection{Investigating the lack of optical emission lines}
Having established, from the X--ray analysis,  the presence  of 
 AGN in the apparently normal galaxies
presented here, we now investigate if the lack of significant 
emission lines in their optical spectra could be explained even 
assuming a standard AGN.
To this end we use a simple model based on an AGN plus early-type galaxy 
optical template\footnote{The noise is not included in our templates.} 
(Francis et al. \cite{Francis}, Elvis et al. \cite{Elvis94};
BC2000\footnote{The Bruzual \& Charlot 2000 
models have been retrieved via anonymous ftp: ftp.iap.fr}) 
to reproduce the available optical spectra. 
The approach used for each source is summarized in the following steps:\\
{\bf 1.} the AGN template is normalized at 2500 \AA~ on the basis 
of the unabsorbed, rest-frame, 2 keV flux using a starting value of 
$\alpha_{ox}$\footnote{The 
$\alpha_{ox}$ is defined here from 2500$\AA$ to 2 keV in the rest frame:
$\alpha_{ox}=-$log(f$_{2500\rm \AA}$/f$_{2\rm keV}$)/log($\nu_{2500\rm \AA}$/$
\nu_{2\rm keV})$.}=1.5 (the typical value of high-luminosity Seyfert and 
quasars, Brandt, Laor, \& Wills \cite{Brandt0});\\
{\bf 2.} assuming a Galactic standard value of 
$E_{B-V}/N_H$=1.7x10$^{-22}$ mag cm$^{-2}$ (Bohlin et al. \cite{Bohlin}) 
and using the intrinsic N$_H$ value derived from the X--ray spectral analysis, 
the 
continuum and the broad line components of the AGN template are absorbed;\\ 
{\bf 3.} the AGN template is redshifted to the z of the source and 
summed with a redshifted early--type galaxy template. 
While the optical flux of the AGN template is then fixed on the
basis of the X--ray properties and the chosen $\alpha_{ox}$, 
the optical normalization of the host galaxy is set 
to reproduce the available spectrum. The latter is  normalized 
to the photometric
data in the R band (see Table~1) or, if possible, to the
fraction of the total light which has passed through the used slit,
taking into account also the atmospheric and seeing conditions 
during the observations;\\
{\bf 4.} if in the final template (AGN+galaxy) evident emission lines are
present (i.e. visible at the S/N ratio of our observed spectrum),
the procedure is restarted from point {\bf 1}, using a
lower value of $\alpha_{ox}$ 
until a good reproduction of the available spectrum
is reached. The resulting $\alpha_{ox}$ represents then an upper limit
on the  $\alpha_{ox}$ of the hidden AGN.

In the next sub--sections we discuss the results obtained for each
of the 3 sources.

\subsection{Source \#1}

\begin{figure*}[!]
{\epsfig{figure=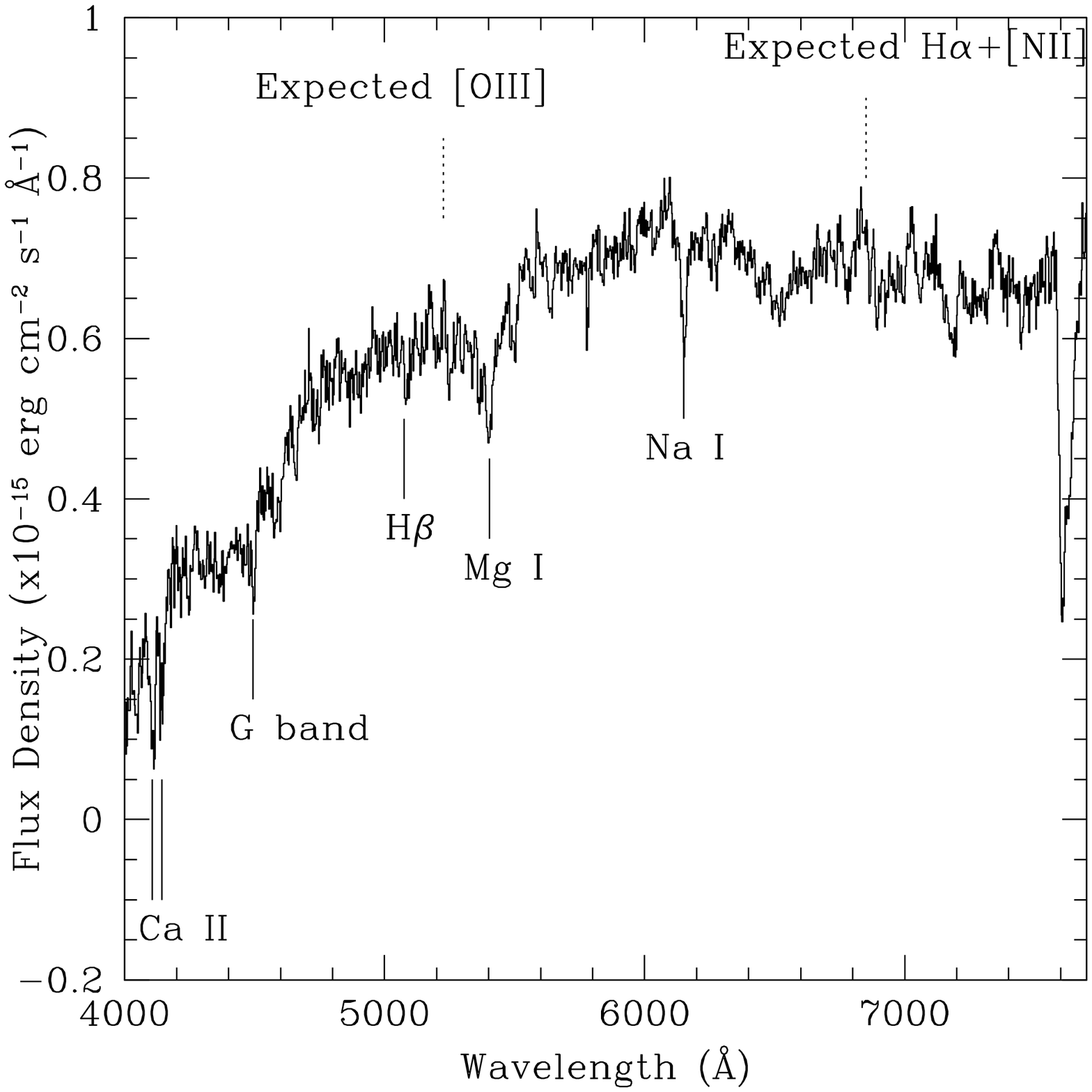, width=8cm, height=8cm}}
{\epsfig{figure=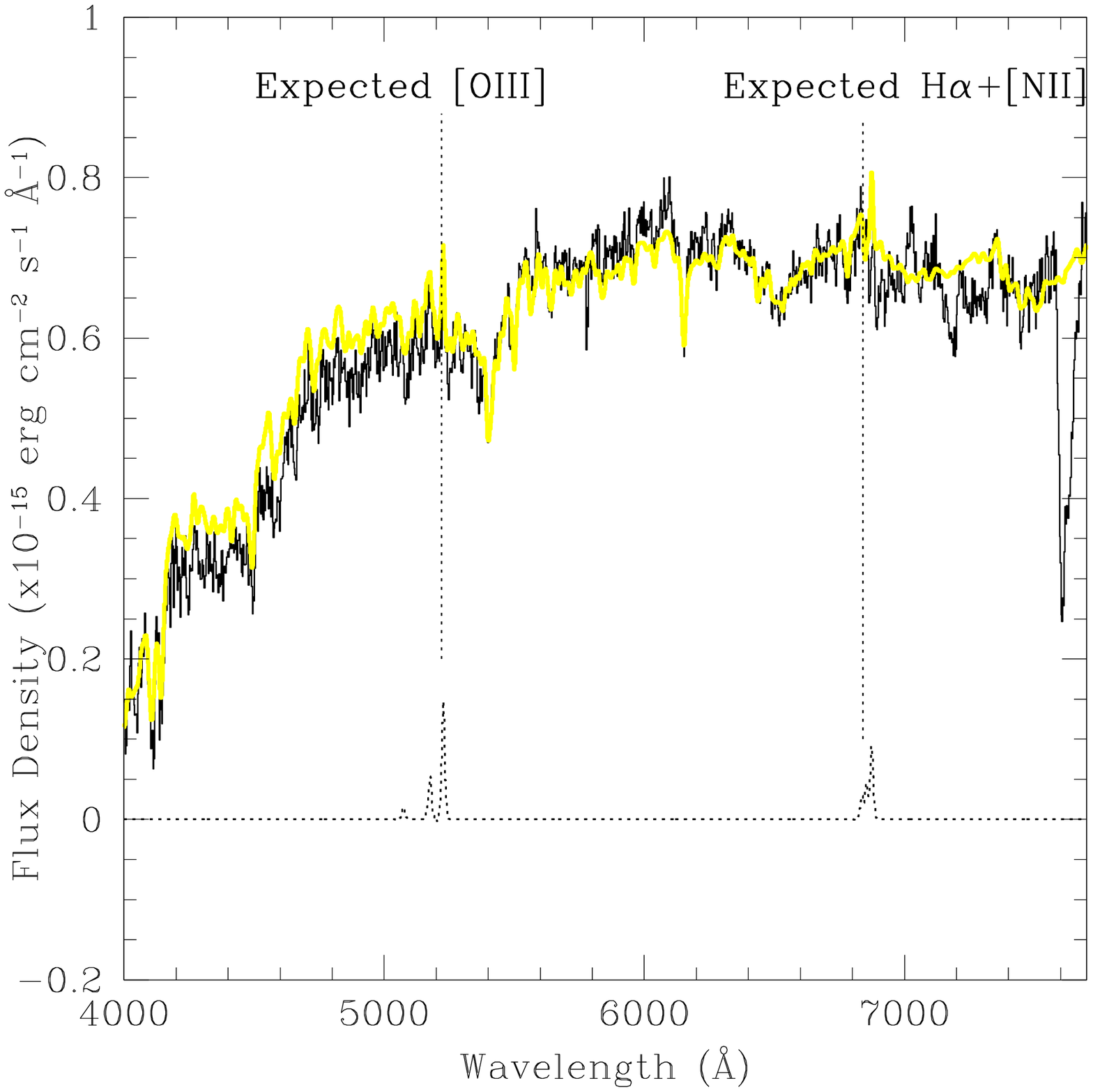, width=8cm, height=8cm}}
\caption{{\it Left panel:} 
 Optical spectrum of XMMJ021822.3-050615.7 taken at 
the 3.58~m TNG telescope.
{\it Right panel:} The optical template (grey solid line), 
obtained by summing a  12~Gyr old early--type galaxy plus an AGN template 
(black dotted line, $\alpha_{ox}\sim$1.25, E$_{B-V}$=35 mag), 
is  overlayed to the observed spectrum (black solid line) of source \#1. 
Only the strongest emission lines present in a typical 
AGN spectrum are labeled.}
\label{0218 Spectrum+Model}
\end{figure*}
Source~\#1 has an X--ray spectrum typical of an obscured AGN.
In this case the torus intercepts the line of sight and both 
the central engine and the {\it Broad Line Region} (BLR) should be hidden.
On the basis of the AGN unified model (Antonucci et al. \cite{Antonucci})
the optical spectrum should then be characterized  by narrow 
emission lines produced by the 
{\it Narrow Line Regions} (NLR) placed outside the torus.
However, no significant narrow emission lines are visible in the optical 
spectrum ({\it left panel}, Fig.~5) taken at the 3.58~m {\it Telescopio
Nazionale Galileo} (TNG).
The observed spectrum has been normalized to the
fraction of the total light which has passed through the slit 
used (1.5\arcsec) under the atmospheric and seeing 
conditions (FWHM$\sim$1\arcsec) 
during the observations.
However the discrepancy between the X--ray properties and the
optical spectrum could be only apparent.
Indeed, using the approach described in Sect.~4.1, we find that an AGN
with an intrinsic $\alpha_{ox}<$1.3 and absorbed by an 
E$_{B-V}$=35$\pm$0.1 mag
(consistent with the measured N$_H\sim$2x$10^{23}cm^{-2}$),
could be completely undetectable in the observed optical spectrum.
Actually, in this case no significant emission lines 
are expected to be detected, as it is shown 
in  Fig.~5, {\it right panel}.
Only hints of [OIII] (EW$\sim$1 \AA) 
and H$\alpha$+[NII] (EW$\sim$1 \AA) lines are present  in the
final model. At the signal--to--noise ratio (S/N) reached in the 
observed spectrum,
these two hints of narrow emission lines are compatible with the intensity
of the noise itself and they are not clearly detectable.
In order to obtain a detection at about 10$\sigma$ of these two
lines, a S/N higher than 70 should be reached.
An alternative approach is to use a narrower slit in order to
include a smaller fraction of the starlight of the galaxy and to make 
the nuclear emission lines detectable.
\begin{figure*}[]
{\centerline{\epsfig{file=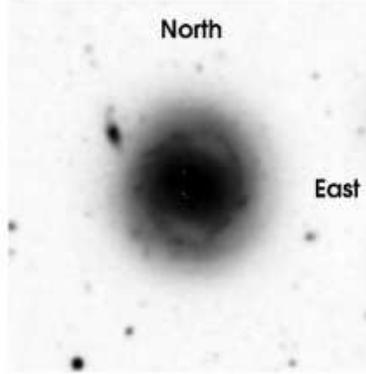, width=10cm, height=8cm, angle=0}}}
\caption{The R-band image ($\sim1\arcmin \times 1\arcmin$) 
of the galaxy obtained with the
{\it Subaru Prime Focus Camera} (Suprime-Cam) of source 
XMMJ021822.3-050615.7.}
\label{}
\end{figure*}
\begin{figure*}[]
{\centerline{\epsfig{file=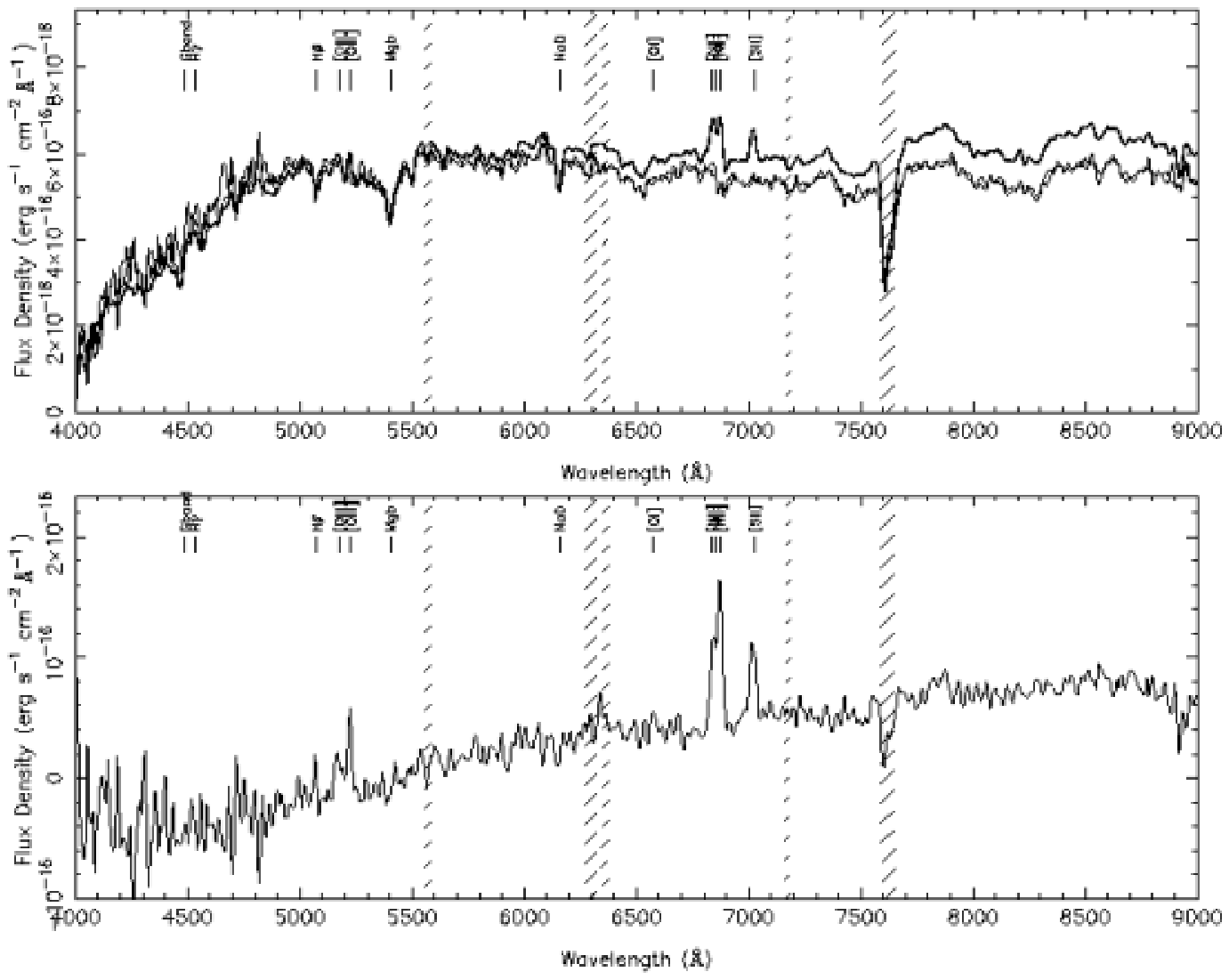, width=10cm, height=15cm, angle=-90}}}
\caption{Optical spectra of XMMJ021822.3-050615.7 taken at 
the 8.2~m Subaru Telescope. 
{\it Top panel:} Optical spectra extracted from the center of 
the source (total spectrum, upper line) and 
from the north and south of the
nuclear region (host galaxy spectrum, lower lines).
The host galaxy spectra are normalized to the total spectrum at 5000A.
{\it Bottom panel:} Total -- Host spectrum.}
\label{}
\end{figure*}
New imaging and spectroscopic observations for this source 
have been recently obtained with the  8.2~m Subaru Telescope.
In Fig.~6 the R--band image of the galaxy obtained with the
{\it Subaru Prime Focus Camera} (Suprime-Cam) is shown.
The source looks like an early type barred galaxy with a weak nucleus and with
a Magellanic Cloud like companion galaxy in the North-West direction.
The spectroscopic observations, carried out with  
the {\it Faint Object Camera And Spectrograph} (FOCAS), 
has been performed using a 0.8\arcsec~slit 
under 0.6\arcsec~seeing conditions. 
The optical spectra obtained for the 
central region (total spectrum) and the host galaxy continuum
are shown in the {\it upper panel} of Fig.~7. 
The host galaxy spectra have been extracted 
from north and south of the nuclear region and then 
normalized to the total spectrum at 5000\AA.
Weak emission lines can be seen in the total
spectrum.
When the host galaxy continuum is subtracted from the total spectrum
(Total -- Host, {\it bottom panel}), 
Seyfert 2 like emission lines can be clearly seen.
This AGN detection technique (data with an  adequate S/N ratio and a proper 
subtraction of the underlying starlight continuum) 
has been already used by Ho et al. (\cite{Ho}, \cite{Ho1}, \cite{Ho2}) 
to search for Low Luminosity AGN.
The Subaru data have thus allowed us to unveil the nuclear engine also
in the optical band and to set more stringent constraints on its intrinsic
optical--to--X--ray spectral index: $\alpha_{ox}\sim$1.2$\pm$0.1. 
In this case, as in the
following, the errors relevant on the
$\alpha_{ox}(AGN)$ values take into account  the uncertainties on both the
X--ray and the optical properties.
We have estimated the ratio between the observed 2-10 keV flux and 
the strength of the [OIII]$_{\lambda5007}$ 
narrow emission line. We have derived a F$_{2-10 keV}$/F$_{[OIII]}$
of 195$\pm$160, 
which is consistent with the value expected for Seyfert galaxies
(Bassani et al. \cite{Bassani}). Moreover, 
this ratio combined with the EW of the measured Fe emission line ($\sim$ 300 
eV) clearly locates this source in the region populated by Compton--thin
sources (see Fig.~1 in Bassani et al. \cite{Bassani}), 
in agreement with our X--ray best fit model.

 In summary, we conclude that a standard AGN, obscured by
N$_H$$\sim$2x10$^{23}$ cm$^{-2}$ and with an intrinsic 2--10 keV
luminosity of about 5x10$^{42}$ erg s$^{-1}$,
can be completely misidentified using only optical spectroscopic
criteria if it is hosted by an early--type galaxy of M$_R\lappr-$22 mag (the 
M$^{*}$ value for a normal galaxy is about -21 mag,
Brown et al. \cite{Brown}).
In particular, the AGN could be completely overwhelmed by the starlight of the
galaxy if it is observed using too wide slits (which include more than 20\%
of the flux of the galaxy) and/or
with a seeing and atmospheric conditions not good enough to reach an adequate 
S/N. It is worth noting that for more distant galaxies ($z>$0.4), 
that typically
populate the deep {\it Chandra} and {\it XMM-Newton} surveys,
it is almost impossible to get an optical spectrum that 
doesn't include more than 20\% of the flux of the host galaxy by using
ground-based telescopes. 
At the moment, the presence of  AGN in object 
like source~\#1 found in deep X--ray surveys can be unveiled in
the optical band only using instrumentations on--board  the
{\it Hubble Space Telescope}.

\subsection{Source \#2 and \#3}
\begin{figure*}[]
{\hskip 0truecm\epsfig{figure=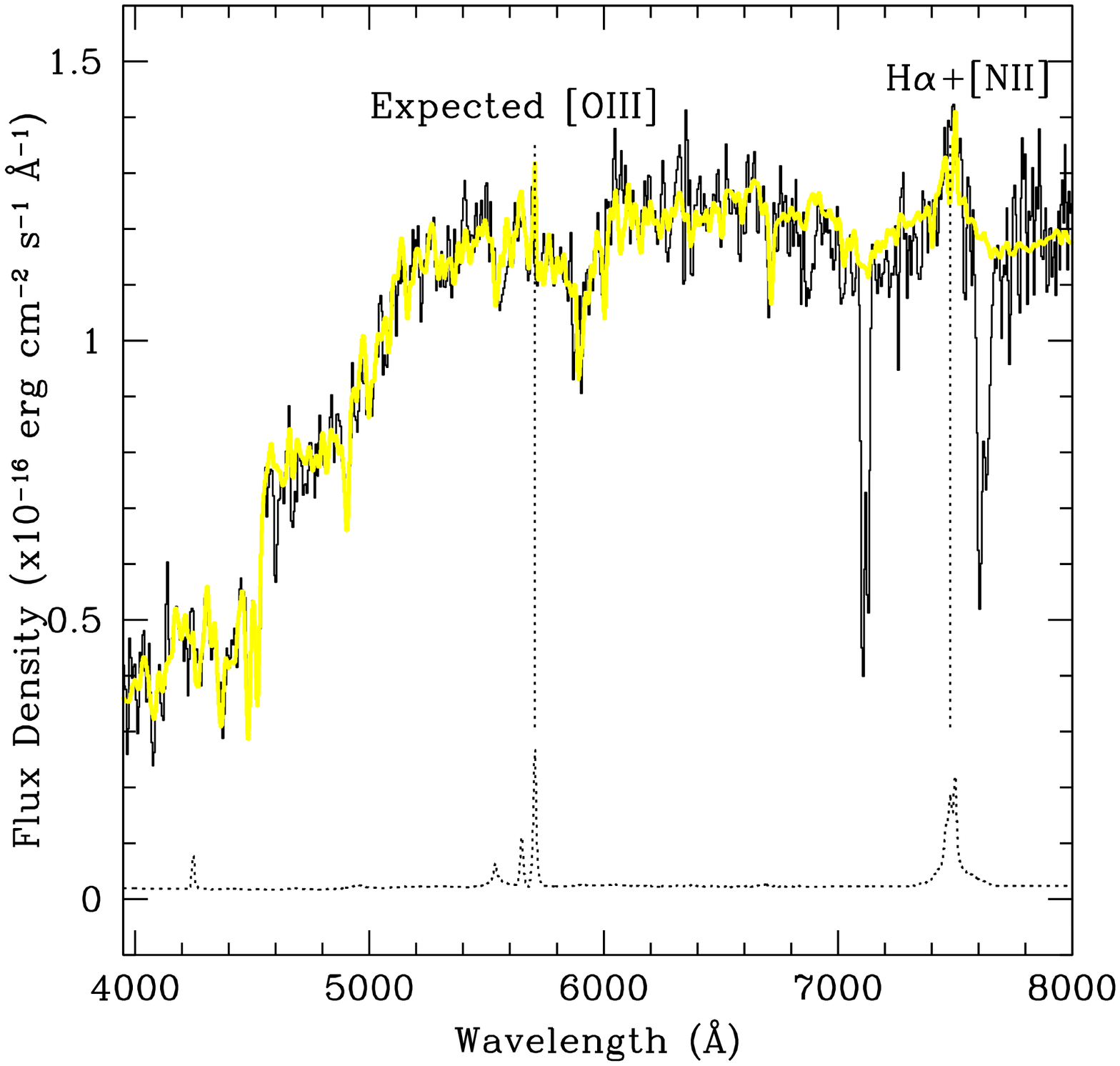, width=8cm, height=8cm}}
{\epsfig{figure=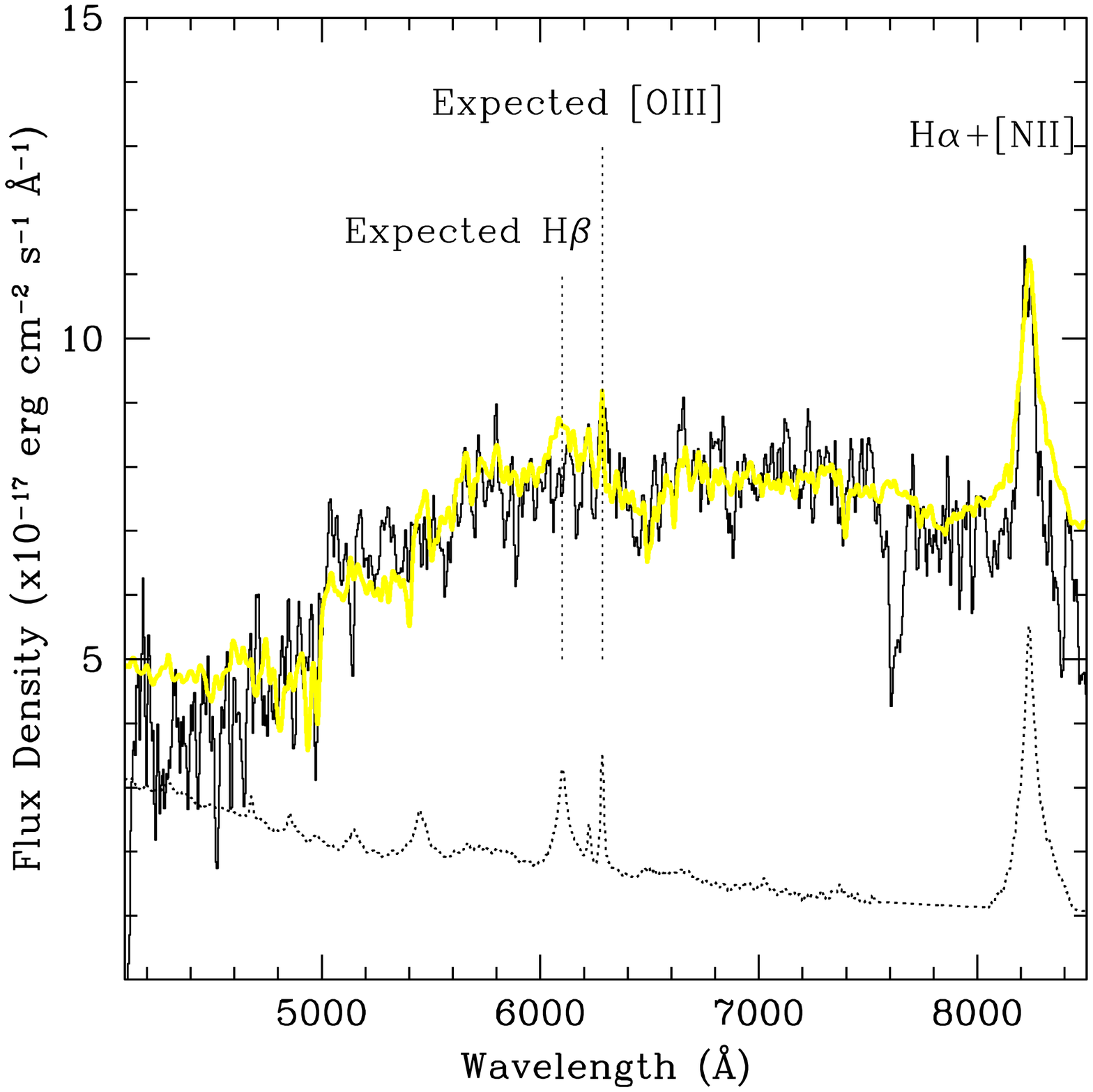, width=8cm, height=8cm}}
\caption{{\it Left panel:} 
 Optical spectrum of XMMJ031859.2-441627.6 (source \#2) taken at 
the 3.6~m ESO telescope (black solid line). The optical template 
(grey solid line), obtained by summing a  5.5~Gyr old early--type galaxy 
plus a AGN template (black dotted line, $\alpha_{ox}\sim$1.2, 
E$_{B-V}$=0.7 mag), is  overlayed to the observed spectrum.
{\it Right panel:} 
Optical spectrum of XMMJ075117.9+180856.1 (source \#3) taken at
the UH 88~inch telescope. The grey solid line represents the 
optical template obtained by summing a 5.5~Gyr old early--type galaxy
with an AGN template (black dotted line, $\alpha_{ox}\sim$1.25, 
E$_{B-V}$=0.2 mag).
Only the strongest emission lines present in a typical 
AGN spectrum are labeled.}
\label{0751,0318 Spectrum+Model}
\end{figure*}
Sources~\#2 and \#3  are respectively slightly obscured 
(N$_H\sim$4x10$^{21}$ cm$^{-2}$) and unobscured 
(N$_H\sim$10$^{21}$ cm$^{-2}$) in the X--ray domain 
and they should appear in the optical band
as intermediate\footnote{Seyfert galaxies that have permitted lines
with narrow and broad components of comparable intensities.} 
(e.g. Maiolino et al. \cite{Maiolino}) 
and/or broad line Seyfert galaxies.

Source \#2 is located along the direction of the cluster Abell~3112 
($\Delta$v=19068 km/s) and its spectroscopic classification 
was taken from the ``ESO Nearby Abell Cluster Survey'' (ENACS) 
catalogue (Katgert et al. \cite{Katgert}, see 
the relevant footnote of Table~1 for the full web page address).
This object was classified as a normal galaxy with no emission lines
visible in the (unpublished) optical spectrum. The spectral range
covered in the rest--frame was from 3940 to 5740 \AA.

A first optical spectrum for source~\#3 was taken at the 3.58~m 
TNG Telescope as part of the AXIS project, covering 
a spectral rest--frame range from 3200 to 6400 \AA.
No evident emission lines were detected in this source and 
it was classified as an optically normal galaxy on the basis of the 
available optical spectrum.

For these two sources (\#2 and \#3), AGN templates have 
been produced using the 
approach described in Sect.~4.1.
Since we do not have enough information 
to normalize the AGN+galaxy template to the amount of light
included in the slit during the observations, 
we have normalized both templates to their  
R band photometric point.
In particular for source \#2, for which we do not have 
in hand the observed optical spectrum,
the type of the host galaxy template has been chosen 
so that the final composite template would match the photometric data in the
B and R bands (Maddox et al. \cite{Maddox}; Katgert et al. \cite{Katgert}).
By assuming an intrinsic $\alpha_{ox}(AGN)<$1.5, we find that, 
although no evident emission lines are expected in the 
observed  ranges, in both sources a prominent H$\alpha$ emission line could 
be present in the yet unsampled region of the spectrum.
For this reason, the two objects have been re--observed 
using a wider spectral coverage. In particular, source \#2 has been
re-observed with the 3.6~m ESO telescope 
(slit of 1.2\arcsec~ and a seeing$\sim$1.6\arcsec),
while new observations for source \#3 have been carried out 
at the UH 88~inches telescope (slit of 1.6\arcsec~ and a seeing$\sim$1.5\arcsec).
In both spectra an evident H$\alpha$ line emerges, 
unveiling the presence of an AGN
in the optical band.
The two new spectra are shown in Fig.~8 (black lines). 
They are normalized on the 
basis of the light included in the slit during the observations.
Thanks to these new observations and using our model, 
an estimate of $\alpha_{ox}$ can be obtained:
$\alpha_{ox}\sim$1.25$\pm$0.2 and $\alpha_{ox}\sim$1.2$\pm$0.2 for
sources \#2 and \#3 respectively.
In Fig.~8 the templates obtained are overlayed (grey lines) to the observed 
spectra.

Also for these two sources, the F$_{2-10 keV}$/F$_{[OIII]}$ ratios
(118$\pm$113 and 410$\pm$400 for sources \#2 and \#3 respectively) are 
consistent with the values expected for Seyfert galaxies and 
the EW$_{\rm Fe-K\alpha}$ upper limits 
(EW$_{\rm Fe-K\alpha}<$ 600 eV for source \#2 and 
EW$_{\rm Fe-K\alpha}<$250 eV for source \#3) indicate that 
these sources are located in the region populated by Compton--thin AGN 
(Bassani et al. \cite{Bassani}), in agreement with the X--ray spectral 
analysis presented here.

On the basis of the $\alpha_{ox}$ quoted above, we have 
estimated the optical magnitudes of the nuclear engine and of the host galaxy:
M$_R$(AGN)$\sim-$18 mag, M$_R$(host)$\sim-$22.5 mag for  source \#2 and  
M$_R$(AGN)$\sim-$21 mag, M$_R$(host)$\sim-$22 mag for  source \#3.

It is worth noting that while for objects like source~\#3 
optical spectroscopy with an appropriate spectral coverage could easily
allow us to recognize the presence of an AGN,
this is not always the case. For instance, in the case of an object
like source \#2, but at a redshift greater than 0.2, 
the fraction of starlight included in the slit  (e.g. 1.2\arcsec) is
usually large enough to  overwhelm the strongest lines, including H$\alpha$.
Using our templates we have estimated that  in such  case (z$>$0.2) the
EW(H$\alpha$) would be $\le$8 \AA, i.e. the H$\alpha$ line is
extremely difficult to detect unless  a very good S/N ratio
is achieved.

\section{Summary and Conclusions}

In this paper we have presented the X--ray and optical
spectra of 3 XMM--BSS
sources selected using the criteria usually adopted to 
define the {\it X-ray Bright Optically Normal Galaxies}.
The good statistics available for the {\it XMM-Newton} data has allowed 
us to unambiguously unveil the presence of an AGN  in all of them.
Once more, this result highlights how the X--ray data are the best
way of finding AGN even for this kind of sources.
In particular, we find that both X--ray obscured 
(N$_H>$10$^{22}$ cm$^{-2}$) and unobscured (N$_H<$10$^{22}$ cm$^{-2}$) AGN,
 covering a range of luminosities from
10$^{42}$ to 10$^{43}$ erg s$^{-1}$, are hosted in these sources.
No Compton--thick objects have been found.

Using the X--ray properties derived for each of the 3 sources, 
a simple model (developed  in order to investigate the apparent
lack of optical emission lines) and new optical
observations, we find that:
\begin{description}
\item --  the AGN considered here are standard AGN 
hosted by galaxies with
a magnitude M$_R$ brighter than M$^*$;
\item -- X--ray obscured AGN with an intrinsic 
L$_{2-10\rm keV}\sim$5$\times$10$^{42}$ 
erg s$^{-1}$ could be hidden in the optical band if they are hosted by 
a galaxy brighter than M$_R\le-$22 mag; 
\item -- up to z$\sim$0.2 slightly X--ray obscured AGN of
L$_{2-10\rm keV}\sim$9x10$^{42}$ erg s$^{-1}$ could be partially
overwhelmed by a host galaxy with M$_R\le-$22.5 mag.
The sampling of the
H$\alpha$ region should  allow the AGN detection in the 
optical band.
For z$>$0.2 also the H$\alpha$ line could be overwhelmed by the
galaxy continuum using ground-based telescope observations;
\item -- X--ray unabsorbed AGN of L$_{2-10\rm keV}\sim$3$\times$10$^{43}$ 
erg s$^{-1}$ could be optically hidden if they are hosted by a galaxy
with M$_R\le-$22 mag and if the spectral coverage doesn't
include the H$\alpha$ line.
\end{description}

In summary, we find that the X--ray and optical properties of the 
sources studied here do not require a non--standard AGN.
The lack of significant emission lines in the
optical spectra could be explained
by an adequate combination of the absorption associated to the AGN,  of
the optical faintness of this latter with respect to the host galaxy and
of an inadequate set--up and atmospheric conditions during the spectroscopic
observations.
In particular, the results reported here clearly show that in 
some cases a non appropriate
wavelength coverage  of the optical spectrum could be the cause of
a misclassification, as an optically normal galaxy, of an X--ray
source which is truly an AGN. 
A similar result was recently presented by Moran et al. (\cite{Moran}) using 
a sample of nearby Seyfert~2 galaxies.
In this respect, it is worth noting that from
the X--ray deep surveys performed with {\it Chandra} 
(Barger et al. \cite{Barger1}, \cite{Barger2}, \cite{Barger3}; 
Hornschemeier \cite{Hornschemeier}) about 40--60\% 
of the identified sources are optically 
classified as galaxies and for $\sim$70\% of them 
the observed range does not include the H$\alpha$ line. 
Our results, thus, remark the importance of being really careful in the 
optical identification of faint X--ray sources for which no good 
optical spectra are available. Moreover, even if the most likely explanation
for the nature of XBONG galaxies proposed by different groups working on
deeper surveys (e.g. Barger et al. \cite{Barger3}; 
Comastri et al. \cite{Comastri1}, \cite{Comastri2}, \cite{Comastri3}) 
is the presence of heavily obscured AGN, our
results suggest that XBONG galaxies could harbor X--ray unobscured AGN
as well. Observational evidence supporting this last suggestion has already 
been presented from deep surveys by Page et al. (\cite{Page}).

%                                     Two column figure (place early!)
%______________________________________________ Gamma_1 (lg rho, lg e)
%   \begin{figure*}
%   \centering
   %%%\includegraphics{empty.eps}
   %%%\includegraphics{empty.eps}
   %%%\includegraphics{empty.eps}
%   \caption{}
%              \label{FigGam}%
%    \end{figure*}
%

\begin{acknowledgements}
We are grateful to M. Elvis and M. J. Page for useful comments.
PS acknowledge financial support by the Italian {\it Consorzio Nazionale per 
l'Astronomia e l'Astrofisica} (CNAA). This work has received partial 
financial support  from ASI (I/R/037/01) 
and from the Italian Ministry of University and Scientific and Technological 
Research (MURST) through grant Cofin 00-02-36. 
Part of the data presented in this paper have been accumulated through
the AXIS project (http://www.ifca.unican.es/$\sim$xray/AXIS/).
XB and FJC acknowledge financial support from the Spanish Ministry 
of Science and Technology, under grant AYA 2000-1690.
The NOT and TNG telescopes are operated on the island of La Palma 
by the Nordic Optical Telescope Scientific Association 
and the Centro Galileo Galilei of the INAF, respectively, in the Spanish
Observatorio del Roque de Los Muchachos of the Instituto de Astrof\'\i
sica de Canarias. 
We would like to thank also all of the staff members of the
Subaru, ESO and Mauna-Kea Telescopes for their support during the observations
and development of the instruments.
The work reported herein is based partly on observations
obtained with XMM-Newton, an ESA science mission with instruments and
contributions directly funded by ESA member states and the USA (NASA).
This publication makes use of data products from the 2MASS
Two Micron All Sky Survey, which is a joint project of the University of 
Massachusetts and the Infrared Processing and Analysis Center/California 
Institute of Technology, funded by the National Aeronautics and Space 
Administration and the National Science Foundation.
This research has made use of the NASA/IPAC Extragalactic Database 
(NED) which is operated by the Jet Propulsion Laboratory, California 
Institute of Technology, under contract with the National Aeronautics 
and Space Administration.

\end{acknowledgements}

\end{document}